\documentclass[twocolumn,prb,showkeys]{revtex4}

\usepackage{graphicx}

\begin{document}

\title{Gold, copper, silver and aluminum nanoantennas
to enhance spontaneous emission}

\author{A. Mohammadi}
\affiliation{Department of Physics, Persian Gulf University,
75196, Bushehr, Iran}
\author{V. Sandoghdar}
\author{M. Agio}
\email[]{mario.agio@phys.chem.ethz.ch}
\homepage[]{www.nano-optics.ethz.ch}
\affiliation{Nano-Optics Group, Laboratory of Physical Chemistry,
ETH Zurich, CH-8093, Zurich, Switzerland}

\date{\today}

\begin{abstract}
We compute the decay rates of emitters coupled to spheroidal 
nanoantennas made of gold, copper, silver, and aluminum. The spectral 
position of the localized surface plasmon-polariton resonance, the 
enhancement factors and the quantum efficiency are 
investigated as a function of the aspect ratio, background index and 
the metal composing the nanoantenna. While copper yields results 
similar to gold, silver and aluminum exhibit different performances.
Our results show that with a careful choice of the parameters 
these nanoantennas can enhance emitters ranging
from the UV to the near-IR spectrum.
\end{abstract}


\maketitle

\section{Introduction}

Single molecules, nanocrystals and nanotubes are relevant light 
emitters for fundamental research and
applications.~\cite{bruchez98,uppenbrink99,oconnell02,ossicini03,moerner04,silbey07}
However, many of these systems exhibit a low quantum yield
and often photobleach. The latter issue can be solved by embedding the 
emitter into a matrix, such that reactive elements like oxygen 
cannot interact with the dye.~\cite{pfab04,boiron96}
Regarding the low quantum yield, a possible solution exploits the concept of
radiative decay engineering with microcavities,~\cite{gerard99} photonic 
crystals~\cite{galli06} or metal nanostructures.~\cite{lakowicz01}
It turns out that a faster radiative decay rate also reduces 
photobleaching, because the emitter is in the excited state for a shorter 
time. Even if microcavities and photonic crystals can be as small as a few 
microns, they still occupy a space much larger than the emitter.
Furthermore, they require a well defined geometry, which gives constraints on
the fabrication method and hence on the choice of the material.

Recently, we have experimentally demonstrated that a single gold 
nanoparticle enhances the fluorescence signal of a single  
molecule~\cite{kuehn06,kuehn08} and found quantitative 
agreement with theory.~\cite{ruppin82} Moreover, our 
calculations show that gold nanoparticles with designed shapes
can increase the decay rates by three orders of magnitude.~\cite{rogobete07}
These so-called {\it nanoantennas}~\cite{greffet05} can thus be
used to improve the quantum efficiency of
emitters~\cite{biteen05,mertens06} and
reduce photobleaching~\cite{hale01} with the advantage that they have 
nanoscale dimensions, a simple shape, and a broad
resonance that does not require fine tuning of the structure parameters.
Furthermore, metal nanoparticles can be mass produced and surface 
functionalization allows controlled binding of the emitter.~\cite{zhang07}

\begin{figure}[h]
\centering
\includegraphics[width=6cm]{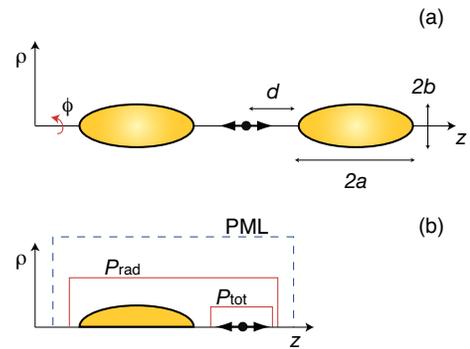}
\caption{\label{layout}A single emitter is coupled to a nanoantenna
made of one or two metal spheroids. 
(a) The dipole is placed at a distance $d$ and
it is oriented along the $z$ axis. The spheroid has dimensions
$a$ and $b$ for the semi-major and semi-minor axes, respectively.
When the nanoantenna consists of two spheroids, the emitter
is at the center of the gap with width $2d$.
The rotational symmetry with respect to the $z$ axis makes the
system a body of revolution that can be treated in two dimensions by 
considering its cross section (b). The total $P_\mathrm{tot}$ and radiated 
$P_\mathrm{rad}$ powers are obtained using Poynting theorem (solid 
lines). The mesh is truncated using PML absorbing boundary conditions 
(dashed line).}
\end{figure}

Nanoantennas base their properties on the so-called 
localized surface plasmon-polariton resonance (LSPR), which is sustained by 
the collective oscillation of free electrons in the metal.
This resonance can be tuned by changing shape, size, 
background index and material.~\cite{bohren83} Because emitters 
cover a broad spectral range, it is interesting to investigate which 
nanoantenna designs should be chosen for operation in a given frequency 
domain. Similar studies have been carried out for the 
field enhancement in surface-enhanced Raman 
scattering.~\cite{moskovits85,cline86,zeman87}

In this paper we study the decay rate enhancement and the quantum efficiency
for an emitter coupled to nanoantennas made of one or two 
spheroids as a function of several 
parameters, including aspect ratio, background index and metal.
We choose spheroidal nanoparticles because they have a simple geometry, 
yet with sufficient degrees of freedom to represent a model system for 
nanoantennas. Indeed, they have been extensively studied
for field-enhanced spectroscopy~\cite{moskovits85,cline86,zeman87}
and for fluorescence enhancement.~\cite{gersten81,rogobete07,agio07,mohammadi08,mertens08}
We discuss nanoantenna designs that cover the spectral range
from the UV to the near-IR. Even if the LSPR
can be easily tuned by changing the spheroid aspect ratio,~\cite{gersten81} 
one has to consider that the decay rates might not be enhanced as much as
desired. Therefore, both geometric effects and
material properties have to be taken into account.

\section{Results and discussion}
\label{results}

\subsection{Theory and computational approach}
\label{computation}

When an emitter is placed in the near field of a nanoantenna,
its radiative decay rate $\gamma_\mathrm{rad}^{o}$ is modified
to $\gamma=\gamma_\mathrm{rad}+\gamma_\mathrm{nrad}$.~\cite{ruppin82}
$\gamma_\mathrm{rad}$ represents the energy that reaches the far
field, while $\gamma_\mathrm{nrad}$ accounts for the radiated energy
absorbed by the nanoantenna due to material losses.
The ratio $\eta_\mathrm{a}=\gamma_\mathrm{rad}/\gamma$ can be considered as a
quantum efficiency. If $\eta_\mathrm{a}$ is small, the emitter is quenched 
even if the radiative decay rate is large~\cite{ruppin82}. 
Another important quantity is the Purcell factor defined as 
$F=\gamma_\mathrm{rad}/\gamma_\mathrm{rad}^{o}$, which represents the
radiative decay rate enhancement. If the isolated emitter has a 
quantum efficiency $\eta_o$, when it is 
coupled to the nanoantenna, it acquires a quantum efficiency $\eta$
that depends on $F$ and $\eta_\mathrm{a}$, which 
reads~\cite{agio07,mohammadi08}
\begin{equation} \label{eta}
\eta=\frac{\eta_o}{(1-\eta_o)/F+\eta_o/\eta_\mathrm{a}}.
\end{equation}
Equation~\ref{eta} shows that if the emitter possesses a poor quantum 
efficiency $\eta_o$, the nanoantenna can effectively enhance it to
a value close to 100\%, if $F\gg 1$ and $\eta_\mathrm{a}\simeq 
100\%$. $F$ and $\eta_\mathrm{a}$ strongly depends on the relative position
and orientation of the emitter with respect to the 
nanoantenna~\cite{ruppin82} and on the nanoantenna shape and 
size.~\cite{rogobete07} Furthermore these quantities depend also on 
the material composing the nanoantenna and on the background medium. For 
simplicity, here we fix the emitter position and orientation and focus on 
the effect of size, shape and material properties.
The emitter is positioned on the nanoantenna axis at a distance d=10 nm
from the spheroid surface and oriented along the spheroid major axis
as shown in Fig.~\ref{layout}(a). For the case of nanoantennas made of two 
spheroids, the emitter is at the center of a 20 nm gap formed between 
the two nanoparticles.

The decay rates are obtained from classical electrodynamics calculations
by collecting the total $P_\mathrm{tot}$ and radiated $P_\mathrm{rad}$ 
powers of an oscillating dipole located at the position of the 
emitter.~\cite{xu00} These quantities are computed using the 
Finite-Difference Time-Domain (FDTD) 
method~.\cite{taflove05,kaminski07} Furthermore, we take advantage of the 
rotational symmetry of the system to reduce the 
problem to two dimensions, see Fig.~\ref{layout}(b), and we employ 
the body-of-revolution FDTD approach.~\cite{taflove05,mohammadi08}
The experimental dielectric function of metals is fitted using Drude or
Drude-Lorentz dispersion models.~\cite{kaminski07,mohammadi08bis}
The FDTD mesh discretization is chosen to be 1 nm for gold and copper 
nanoantennas, while for silver and aluminum nanoantennas we use 0.5 nm
to compensate for the shorter operating wavelength. We terminate
the FDTD mesh with perfectly-matched-layer (PML) absorbing boundary  
conditions.~\cite{prather99}

\begin{figure}
\includegraphics[width=7.5cm]{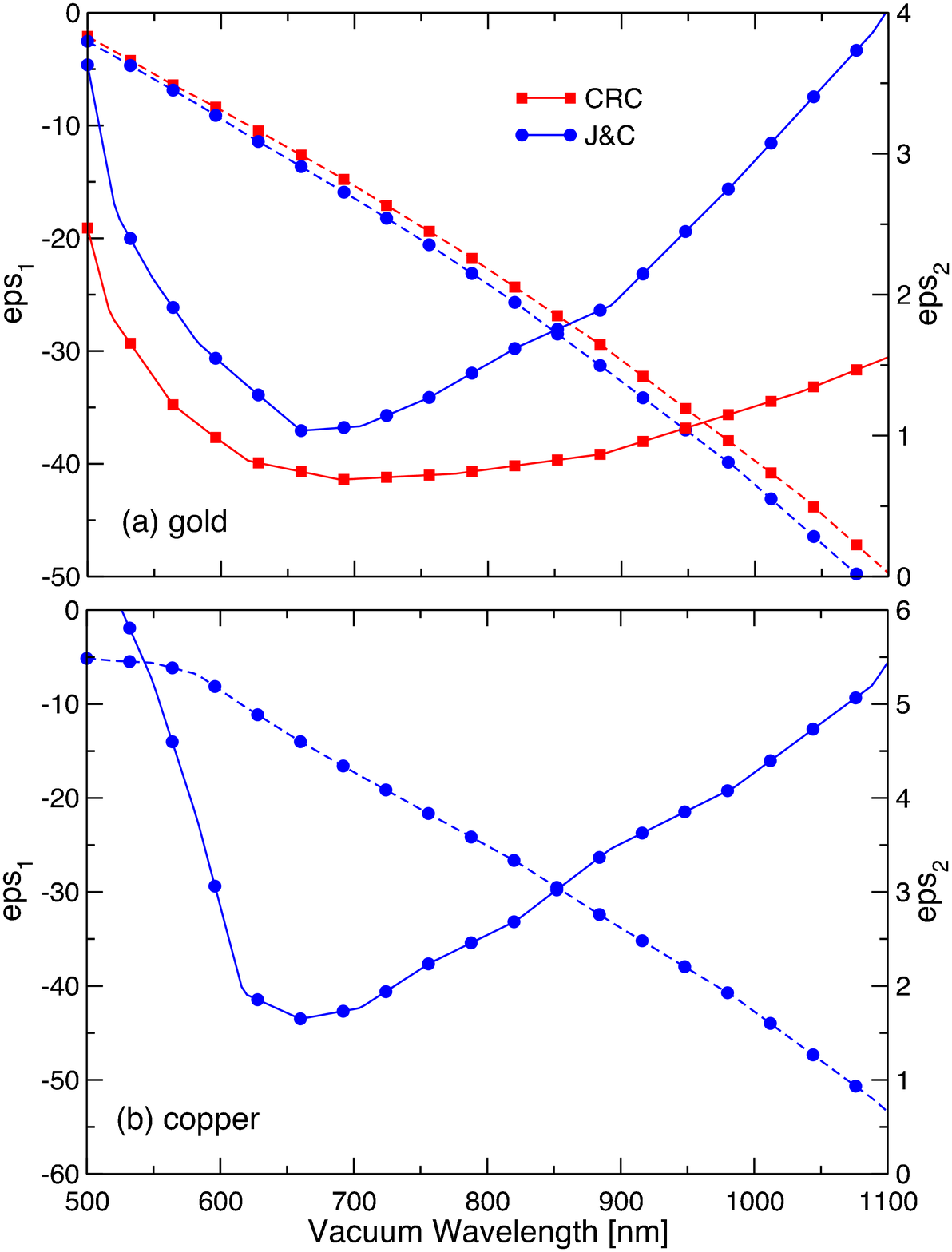}
\caption{\label{eps-Au-Cu}Real (dashed curves) and imaginary parts 
(solid curves) of the dielectric functions of (a) gold and (b) copper. 
The experimental data are compiled from Refs.~\onlinecite{CRChandbook} (CRC) 
and~\onlinecite{johnson72} (J\&C).}
\end{figure}

\subsection{Gold and copper nanospheroids}

To better understand the performances of nanoantennas we first review 
the optical properties of gold and copper. Figure~\ref{eps-Au-Cu} shows
the real and imaginary parts of the dielectric functions of gold and copper
in the visible and near IR spectral range. The real part for the two 
materials is quite similar, whereas the imaginary part for copper is 
slightly larger than for gold if the experimental data are taken from 
Ref.~\onlinecite{johnson72}. Therefore, we expect similar results for 
both materials.
On the other hand, if for gold we consider the experimental values from 
Ref.~\onlinecite{CRChandbook}, the imaginary part gets smaller and 
consequently gold  nanoantennas should further improve 
with respect to copper. We choose the optical 
constants from Ref.~\onlinecite{CRChandbook} for gold and from 
Ref.~\onlinecite{johnson72} for copper.

\begin{figure}
\includegraphics[width=7.5cm]{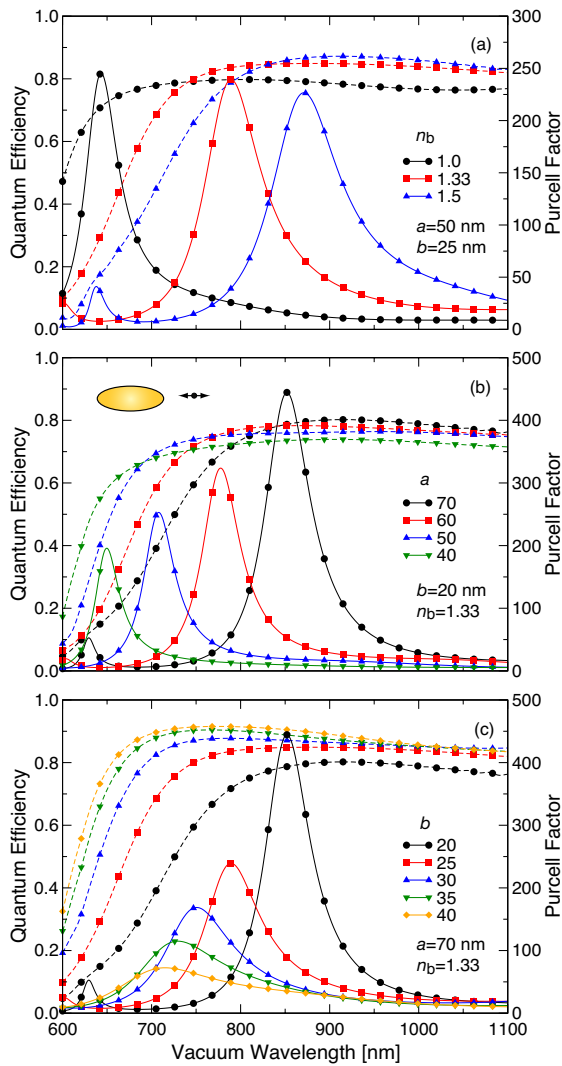}
\caption{\label{1Au}Purcell factor (solid curves) and quantum efficiency 
$\eta_\mathrm{a}$ (dashed curves) for an emitter coupled to a gold spheroid
for $d=10$ nm (see Fig.~\ref{layout}(b)). (a) Dependence on the background 
index $n_\mathrm{b}$ for $a=50$ nm and $b=25$ nm. (b) Dependence on the 
semi-major axis $a$ for $b=20$ nm and $n_\mathrm{b}=1.33$. (c) Dependence 
on the semi-minor axis $b$ for $a=70$ nm and $n_\mathrm{b}=1.33$.}
\end{figure}

Since we have already studied nanoantennas made of two gold 
spheroids,~\cite{mohammadi08} here we focus the attention on 
single ones. This system can be also studied using an approximate
method developed by Gersten and Nitzan~\cite{gersten81} and recently
improved by Mertens et al.~\cite{mertens07,mertens08} to account for
radiative damping~\cite{wokaun82} and depolarization effects.~\cite{meier83}
Figure~\ref{1Au}(a) elucidates how the Purcell factor and the quantum 
efficiency $\eta_\mathrm{a}$ depend on the background index for an emitter 
coupled to
a gold spheroid with semi-axes $a=70$ nm and $b=25$ nm. Even a small 
change in the refractive index shifts the LSPR by more than 
hundred nanometers. At the same time, the resonance gets wider because
radiative broadening increases with the refractive index.~\cite{wokaun82}
That also explains the small decrease in the Purcell factor.
As a consequence of material losses, the quantum efficiency drops to zero 
below 600 nm. However, the shift of the LSPR towards shorter
wavelengths improves the quantum efficiency. For instance, it is 
larger than 70\% around 650 nm if the nanoantenna is embedded in air,
$n_\mathrm{b}=1$. 

\begin{figure}
\includegraphics[width=7.5cm]{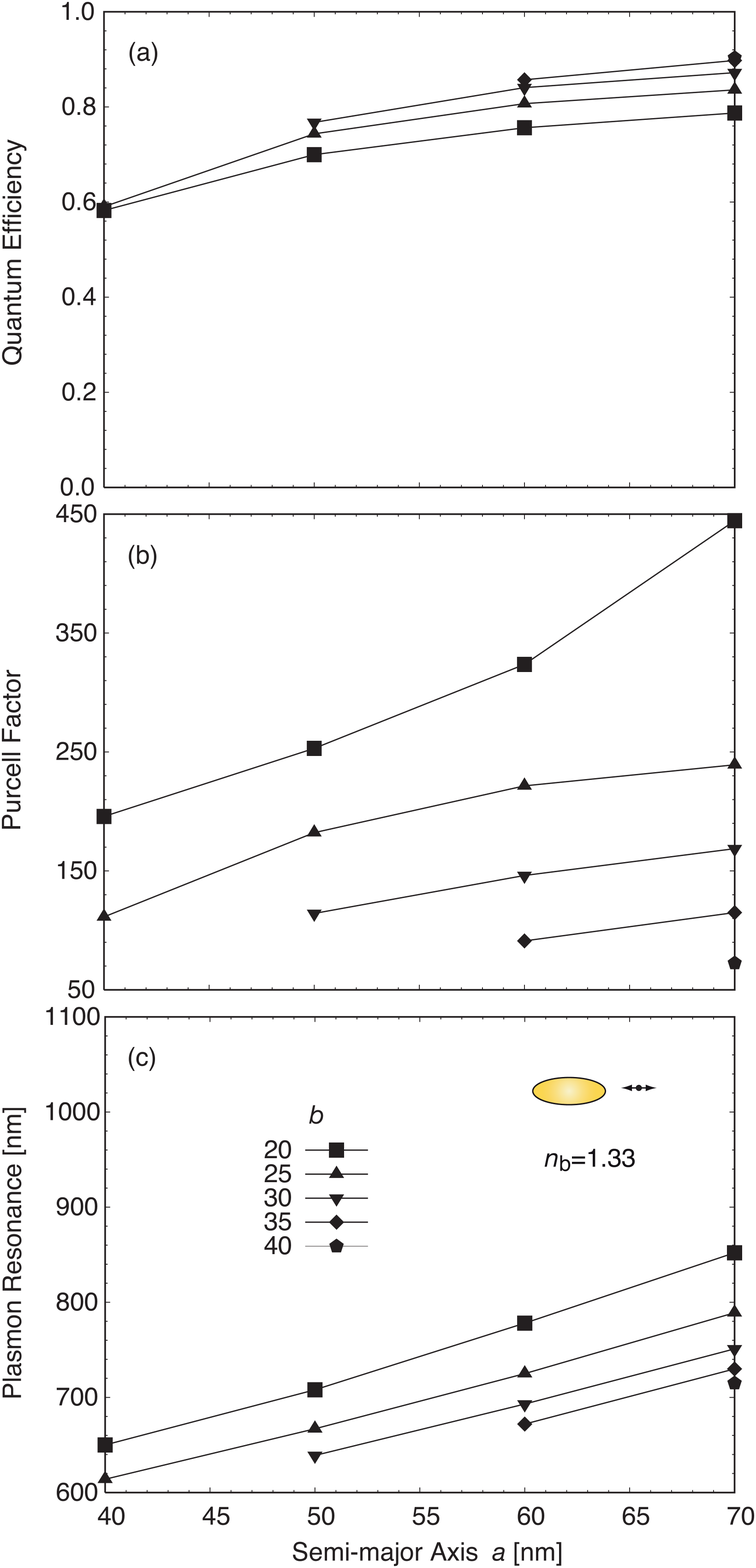}
\caption{\label{1AueB1.7689}(c) LSPR wavelength, corresponding to the 
maximum Purcell factor for an emitter coupled to a gold spheroid, as a 
function of $a$ and $b$ (see Fig.~\ref{layout}(b)). The distance to the 
spheroid is $d=10$ nm and the background index is $n_\mathrm{b}=1.33$. 
(b) Purcell factor and (a) quantum efficiency $\eta_\mathrm{a}$ for the 
corresponding wavelengths and spheroid parameters given in (c).}
\end{figure}

\begin{figure}
\includegraphics[width=7.5cm]{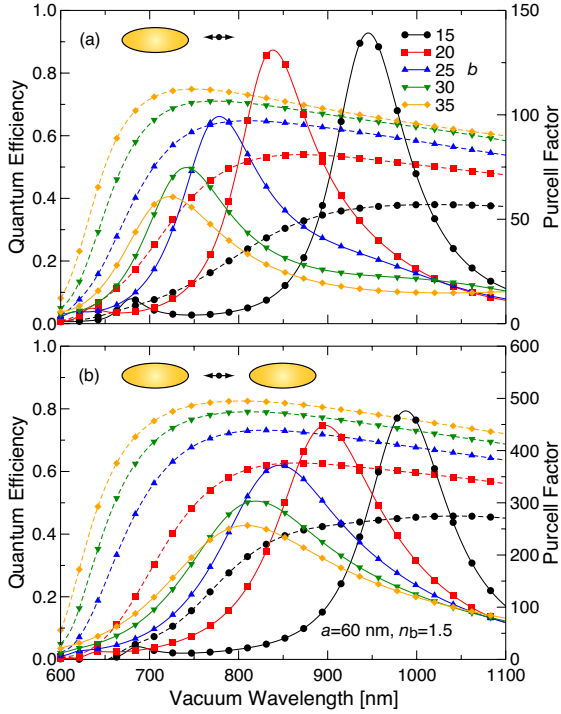}
\caption{\label{12Cu}Purcell factor (solid curves) and quantum efficiency 
$\eta_\mathrm{a}$
(dashed curves) for an emitter coupled to a nanoantenna made of (a) one 
or (b) two copper spheroids in glass, with $a=60$ nm and $d=10$ 
nm (see Fig.~\ref{layout}).}
\end{figure}

Figures~\ref{1Au}(b) and~\ref{1Au}(c) present the situation where the 
background index $n_\mathrm{b}$ is fixed to that of 
water, $n_\mathrm{b}=1.33$, and the spheroid
axes are varied. In Fig.~\ref{1Au}(b) the semi-minor axis is constant,
$b=20$ nm, and the semi-major one spans from 40 to 70 nm. When the aspect 
ratio gets smaller the LSPR shifts towards shorter 
wavelengths and the Purcell factor drops.~\cite{mohammadi08} Notice that 
even if a smaller aspect ratio implies a smaller volume and a dipolar 
LSPR closer to the higher order modes,~\cite{mertens08} the quantum 
efficiency can still be large, as shown for $a=40$ nm. Also decreasing the 
volume reduces the effect of radiative broadening and the LSPRs
appear narrower.
In Fig.~\ref{1Au}(c) we keep the semi-major axis constant, $a=70$ nm, and
vary the spheroid width. In this case, reducing the aspect ratio increases
the volume such that radiative broadening increases and the LSPRs appear 
wider. For the same aspect ratio, the smaller spheroid
in Fig.~\ref{1Au}(b) with $a=40$ nm and $b=20$ nm exhibits a stronger 
Purcell factor and a lower quantum efficiency than the larger one 
with $a=70$ nm and $b=35$ nm.

\begin{figure}
\includegraphics[width=7.5cm]{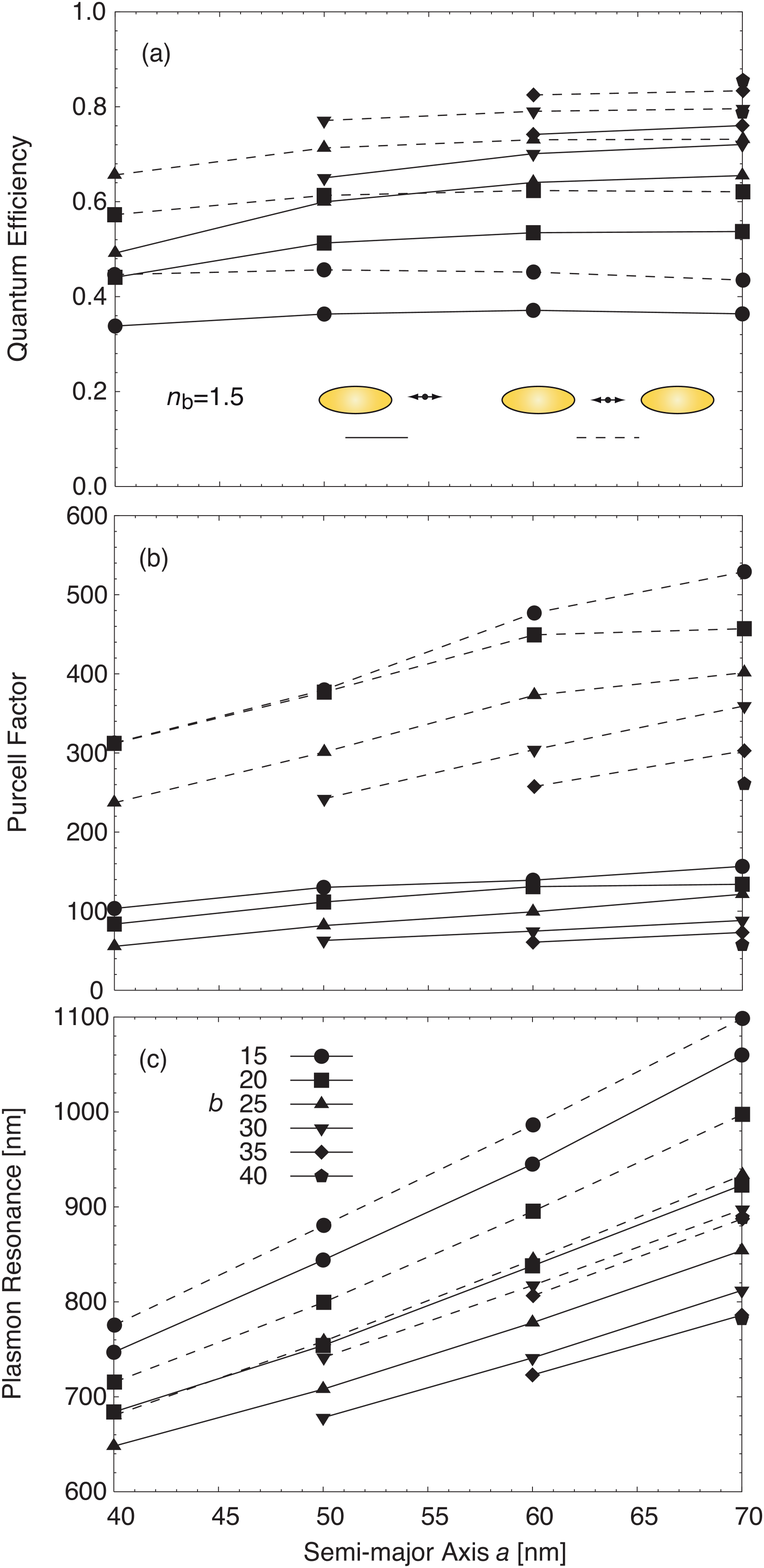}
\caption{\label{CueB2.25}(c) LSPR wavelength, corresponding to the 
maximum Purcell factor for an emitter coupled to one (solid curves), or 
two (dashed curves) copper spheroids, as a function of $a$ and $b$ (see 
Fig.~\ref{layout}). The distance to the 
spheroid is $d=10$ nm and the background index is $n_\mathrm{b}=1.5$. 
(b) Purcell factor and (a) quantum efficiency $\eta_\mathrm{a}$ for the 
corresponding wavelengths and spheroid parameters given in (c).}
\end{figure}

Figure~\ref{1AueB1.7689} summarizes the results of a single gold spheroid
in water for different values of the nanoantenna axes. In 
Fig.~\ref{1AueB1.7689}(c)
we plot the wavelengths at which the maximum Purcell factor is achieved,
corresponding to the peak of the LSPR. These values are 
reported in Fig.~\ref{1AueB1.7689}(b). For the same wavelengths
we have also computed the quantum efficiency $\eta_\mathrm{a}$, shown in 
Fig.~\ref{1AueB1.7689}(a).
The data for nanoantennas with resonances outside the wavelength range
from 600 to 1100 nm have not been considered. While the quantum 
efficiency does not depend much on the spheroid parameters, the Purcell 
factor changes by almost an order of magnitude, as already seen
in Fig.~\ref{1Au}.

We now move our attention to copper spheroids. Figure~\ref{12Cu}(a)
shows the Purcell factor and the quantum efficiency $\eta_\mathrm{a}$ for an 
emitter coupled
to a single spheroid in glass, $n_\mathrm{b}=1.5$, for $a=60$ nm and variable
$b$. Compared to gold (see Fig.~\ref{1Au}(c)) the enhancement is smaller and 
the resonances are broader as expected by the fact that the imaginary 
part of copper is larger. On the other hand, the Purcell factor does not 
drop as rapidly when the aspect ratio decreases.
The quantum efficiency is lower, but it shows the same trend. Namely,
if the LSPR shifts to shorter wavelengths, the efficiency 
increases. For an aspect ratio equal to 2, for $a=60$ nm and $b=30$ nm, the
Purcell factor is about 75 and the quantum efficiency is close to 70\%.
If we consider a nanoantenna made of two copper spheroids, we can improve
both the Purcell factor and the quantum efficiency, but we also redshift
the resonance wavelength, as shown in Fig.~\ref{12Cu}(b).

Collective data on the resonance wavelength, Purcell factor and quantum 
efficiency are displayed in Fig.~\ref{CueB2.25} for nanoantennas made of 
one or two copper spheroids in glass. In 
Fig.~\ref{CueB2.25}(a) notice that the quantum efficiency is now more 
sensitive to the nanoantenna geometry than in the case of gold (see 
Fig.~\ref{1AueB1.7689}(a)), while the opposite holds for the Purcell factor, 
when comparing Fig.~\ref{CueB2.25}(b) and Fig.~\ref{1AueB1.7689}(b).
These differences stem from the imaginary part
of the dielectric function, which is larger for copper.
We should keep in mind that if for gold we used the optical constants from
Ref.~\onlinecite{johnson72}, gold and copper would exhibit even closer 
resemblance.

\begin{figure}
\includegraphics[width=7.5cm]{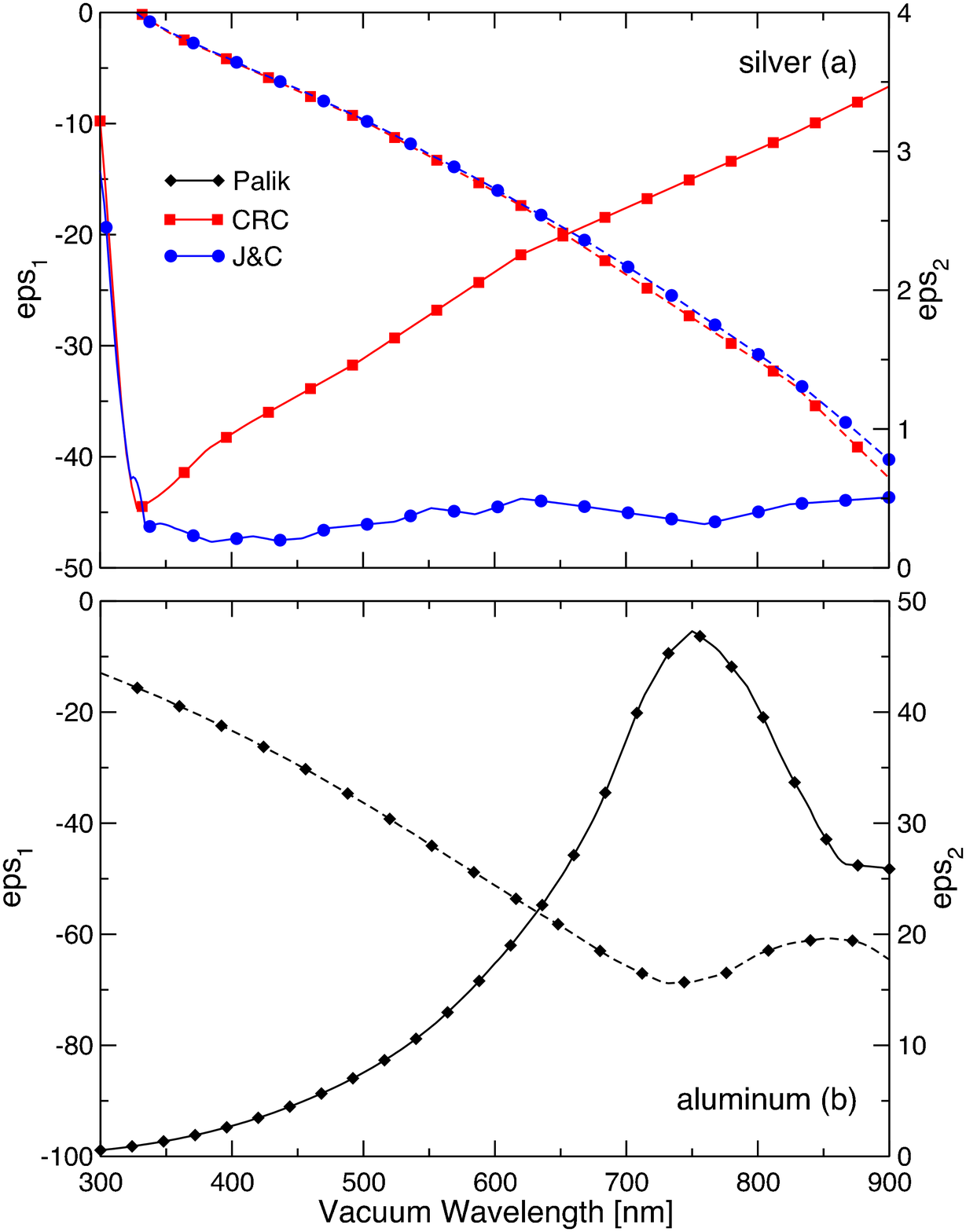}
\caption{\label{eps-Ag-Al}Real (dashed curves) and imaginary parts 
(solid curves) of the dielectric functions of (a) silver and (b) 
aluminum. The experimental data are compiled from 
Refs.~\onlinecite{CRChandbook} (CRC),~\onlinecite{johnson72} (J\&C), 
and~\onlinecite{palik98} (Palik).}
\end{figure}

\subsection{Silver and aluminum nanospheroids}

We now consider nanoantennas made of silver or aluminum. As before, we 
start looking at the real and imaginary parts of the dielectric function, 
presented in Fig.~\ref{eps-Ag-Al}. Silver appears to be similar to gold if 
the experimental data are taken from Ref.~\onlinecite{CRChandbook} and from 
Ref.~\onlinecite{johnson72}, respectively. The main difference is that
silver has a higher plasma frequency so that the curves are shifted
towards shorter wavelengths. Therefore, we expect that silver yields results
similar to gold, but in a spectral range closer to UV light. However, if 
for silver we consider the experimental data of Ref.~\onlinecite{johnson72},
we notice that while the real part is almost the same, the imaginary part
drops to much lower values. In this case, silver nanoantennas should perform 
much better than their gold counterparts. 
Because samples of silver nanoantennas might exhibit
a lower optical quality than the bulk material, caused
by imperfections in the crystalline structure and contamination
occurring in the nanofabrication steps, we prefer to choose 
the experimental dielectric function with the largest imaginary 
part.~\cite{CRChandbook}

\begin{figure}
\includegraphics[width=7.5cm]{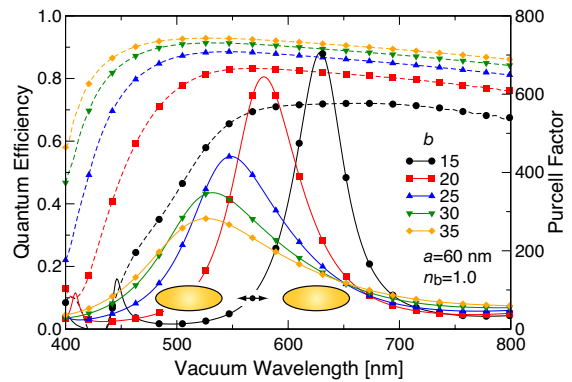}
\caption{\label{12Ag}Purcell factor (solid curves) and quantum efficiency 
$\eta_\mathrm{a}$ (dashed curves) for an emitter coupled to a 
nanoantenna made of two silver spheroids in air with $a=60$ nm and 
$d=10$ nm (see Fig.~\ref{layout}(a)).}
\end{figure}

Figure~\ref{eps-Ag-Al}(b) displays the optical constants of aluminum as given
in Ref.~\onlinecite{palik98}. Aluminum has a plasma frequency even higher 
than silver. Therefore the real part is larger in the same spectral range.
On the other hand, there is an interband absorption peak located at 800 nm,
which creates a dispersive profile in the real part of the dielectric 
function and, most importantly, a strong increase in the imaginary part.
This makes aluminum less attractive for nanoantenna 
applications in the spectral range around 800 nm. Even if the imaginary part
is significantly larger than in the noble metals, in the region below 600 nm
the large and negative real part ensures that the skin depth
is sufficiently small to prevent significant losses.

\begin{figure}
\includegraphics[width=7.5cm]{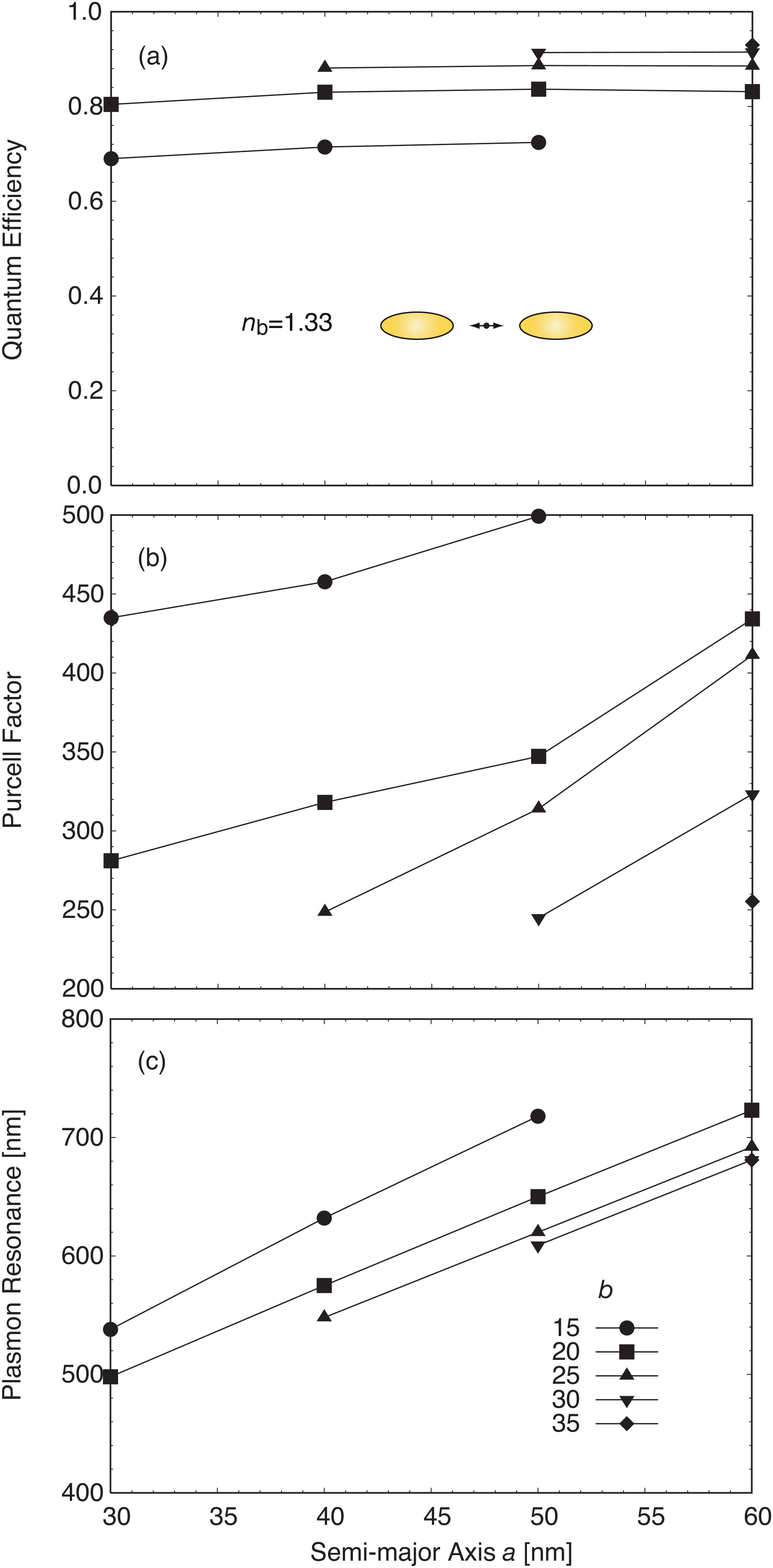}
\caption{\label{2AgeB1.7689}(c) LSPR wavelength, corresponding to the 
maximum Purcell factor for an emitter coupled to two silver spheroids, as 
a function of $a$ and $b$ (see Fig.~\ref{layout}(a)). The distance to the 
spheroids is $d=10$ nm and the background index is $n_\mathrm{b}=1.33$. 
(b) Purcell factor and (a) quantum efficiency $\eta_\mathrm{a}$
for the corresponding wavelengths and spheroid parameters given in (c).}
\end{figure}

Figure~\ref{12Ag} shows the quantum efficiency $\eta_\mathrm{a}$ and the 
Purcell factor for an emitter coupled to a nanoantenna made of two 
silver spheroids in air. The general trend agrees with what we have previously
discussed for gold in Fig.~\ref{1Au}(c), and for copper in
Fig.~\ref{12Cu}(b). Because the plasma frequency of silver is higher than 
that of gold and copper, the resonances are shifted by about 200 nm towards
shorter wavelengths. Furthermore, the quantum efficiency and the Purcell 
factor are higher. Using the optical constants of Ref.~\onlinecite{johnson72}
would have yielded even better results. A more complete set of results 
for silver nanoantennas embedded in water is given
in Fig.~\ref{2AgeB1.7689}. In comparison to Fig.~\ref{12Ag}, the LSPR
is redshifted and the Purcell factor is slightly reduced due to
the radiative broadening effect as seen for gold in Fig.~\ref{1Au}(a).

\begin{figure}
\includegraphics[width=7.5cm]{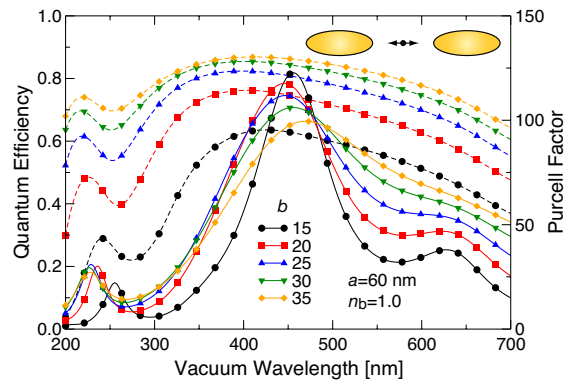}
\caption{\label{2Al}Purcell factor (solid curves) and quantum efficiency 
$\eta_\mathrm{a}$ (dashed curves) for an emitter coupled to a 
nanoantenna made of two aluminum spheroids in air, with 
$a=60$ nm and $d=10$ nm (see Fig.~\ref{layout}(a)).}
\end{figure}

The quantum efficiency $\eta_\mathrm{a}$ and the Purcell factor for an 
emitter coupled to a nanoantenna made of two aluminum spheroids in 
air is provided in Fig.~\ref{2Al}. While the quantum efficiency, as 
expected, increases with the volume of the spheroid, the LSPR
is not redshifted when the aspect ratio increases. The reason for that
can be found in the electromagnetic interaction between the two spheroids.
For a single aluminum spheroid, the LSPR exhibits a small 
redshift in agreement with the polarizability 
theory.~\cite{meier83,zeman87} For the case of two
aluminum spheroids separated by a gap $2d=20$ nm, the interaction between
the two LSPR modes is stronger for small aspect ratios than for larger 
ones because sharper particles have larger but more rapidly decaying 
near fields at their tips. The coupling between the two particles 
redshifts the LSPR.~\cite{aravind81} The increased interaction 
explains also why the Purcell factor does not drop much when the aspect 
ratio decreases: the two spheroids act together more effectively to 
increase the near field. An indication of the same effect can be 
appreciated also for copper in Fig.~\ref{12Cu}, where however the 
redshift caused by the single 
particle polarizability is so strong that makes it difficult to notice.
The Purcell factors given by the aluminum nanoantennas of Fig.~\ref{2Al} are
not as large as found for the same system made from other materials.
Because the quantum efficiency is large, the reason for that should be mainly
attributed to radiative broadening rather than to losses.~\cite{wokaun82}
For instance, since the radiative broadening is proportional to 
$1/\lambda^3$, the effect is 8 times stronger at 400 nm than at 800 nm.
Indeed calculations of the field enhancement have shown that the LSPR
should be located around 200-300 nm and the semi-major axis of the
spheroid should not be larger than 40 nm for optimal 
performances.~\cite{zeman87} Therefore aluminum nanoantennas can be better
exploited for the UV spectral region rather than for the visible and near IR 
range.

\section{Conclusions}
\label{conclusion}

We have investigated the performances of nanoantennas for improving light 
emitters by considering different materials, namely gold, copper, silver 
and aluminum, aspect ratios and background media. While gold and copper 
can both operate in the near IR spectral range, silver is more suitable 
for the visible range and aluminum for the UV range. Therefore, various 
emitters can be enhanced by choosing appropriate nanoantenna parameters.

We have seen that contrary to conventional antennas, nanoantennas
cannot be simply scaled to operate at different wavelengths. Here the
material properties play a fundamental role.
Also the choice of the experimentally determined optical constants
available in the literature can be an issue of concern.~\cite{stoller06}
In particular, these data have been obtained for bulk samples, while 
nanoantennas are truly nanoscale objects. Even if the 
volume of a nanoantenna is sufficiently large to ignore quantum-size 
effects,~\cite{bohren83} the fabrication methods might influence the
actual optical properties by nanograins formation and material 
contamination.~\cite{johnson72,shinya97}

\section*{Acknowledgments}
We thank L. Rogobete, F. Kaminski, Y. Ekinci, N. Mojarad, H. Eghlidi, and 
S. G\"otzinger for helpful discussions. A. Mohammadi is thankful to the
Persian Gulf University Research Council for partial support.
This work was financed by the ETH Zurich initiative on Composite Doped 
Metamaterials (CDM).

\end{document}